\documentclass[prb,twocolumn]{revtex4}
\usepackage{bm}
\usepackage{graphicx}
\usepackage{amssymb}
\usepackage{amsmath}
\usepackage{eufrak}
\usepackage{color}
\usepackage[utf8]{inputenc} 
\usepackage{hyperref}
\usepackage{pifont}
\usepackage{ulem}
\usepackage{array}
\usepackage{verbatim} 
\usepackage{placeins}
\usepackage{tabularx}
	\newcolumntype{C}[1]{>{\centering\arraybackslash}p{#1}} 
\newcommand{\nix}[1]{}

\makeatletter
\newcommand{\bibaddress}[1][]{\item[\@biblabel{#1}]}
\makeatother
\begin{document}	
	\title{Infrared/Terahertz Spectra of the Photogalvanic Effect
		\\  in (Bi,Sb)Te based Three Dimensional Topological Insulators 	}
	
	\author{
		H.~Plank$^1$, J.~Pernul$^1$, S.~Gebert$^1$, S.\,N.~Danilov$^1$, J.~K\"{o}nig-Otto$^2$, 
		S.~Winnerl$^2$,  M.~Lanius$^3$, J.~Kampmeier$^3$, G.~Mussler$^3$, I.~Aguilera$^4$, 
		D.~Gr\"{u}tzmacher$^3$, and S.\,D.~Ganichev$^1$	}
	\affiliation{$^1$Terahertz Center, University of Regensburg, 93051 Regensburg, Germany}
	\affiliation{$^2$Helmholtz Zentrum Dresden, 01328 Rossendorf, Germany}
	\affiliation{$^3$Peter Gr\"unberg Institute (PGI-9) \& J\"ulich Aachen Research Alliance (JARA-FIT), 52425 J\"ulich, Germany}
	%
	\affiliation{$^4$Peter Gr\"unberg Institute (PGI-1) \&  Institute for Advanced Simulation (IAS-1), Forschungszentrum J\"ulich and JARA,  52425 J\"ulich, Germany}
	%
\begin{abstract}	
We report on the systematic study of infrared/terahertz spectra of photocurrents in (Bi,Sb)Te based three dimensional topological insulators. 
We demonstrate that in a wide range of frequencies, ranging from fractions up to tens of terahertz, the photocurrent is caused by the linear photogalvanic effect (LPGE) excited in the surface states.
The photocurrent spectra reveal that at low frequencies the LPGE emerges due to free carrier Drude-like absorption. The spectra allow to determine the room temperature carrier mobilities in the surface states despite the presents of thermally activate residual impurities in the material bulk.  
In a number of samples we observed an enhancement of the linear photogalvanic effect at frequencies between $30\div 60$~THz, which is attributed to the excitation of electrons from helical surface to bulk conduction band states. 
Under this condition and applying oblique incidence we also observed the circular photogalvanic effect driven by the radiation helicity. 
\end{abstract}
\maketitle{} 
\section{Introduction}
Three dimensional topological insulators (TIs) caught attention soon after their prediction, for reviews see~[\onlinecite{Hasan,Qi,Ando}]. 
The band structure at the surface is the reason for their unique features: alike to graphene~\cite{Neto2009}, 
the surface states of topological insulators are characterized by a linear energy dispersion, which is described by the zero mass Dirac equation. The single Dirac cone in TIs, however, leads to a spin-momentum locking and, with that, to new physics. Renowned techniques such as 
angle resolved photoemission spectroscopy (ARPES)~\cite{Hasan, Qi, Ando, ARPES_1} 
or magneto-transport measurements, for review see~[\onlinecite{Bardarson2013}],
are applied  to access and characterize surface carriers in TIs. New opportunities to study Dirac 
fermions are offered by nonlinear high frequency transport phenomena~\cite{GlazovGanichev_review}
which scale with the second or third power of radiation electric field. 
A plethora of such effects has been theoretically discussed and observed in TIs systems including circular and linear photogalvanic effects in three dimensional (3D) TIs~\cite{HosurBerry, Sheng2012, McIver2012, Olbrich2014, Duan2014,Junck2014, Dantscher2015, Hamh2016, Okada2016, Plank2016, Plank2016_2, Pan2017}, edge photogalvanics in two dimensional (2D) TIs~\cite{TItheory, Magarill2016, Magarill2016_2, Dantscher2017}, quantum interference controlled photocurrents~\cite{Bas2015, Bas2016}, ultrafast photocurrents in TI states~\cite{Kastl2012, Kastl2015, Braun2016, Kastl2017, Seifert2017} transient photocurrents in the topological surface state measured by ARPES and its modifications~\cite{2ppe3, Shikin2016, 2ppe4, Shikin2017}, inverse spin-galvanic effect~\cite{inverseTI}, and harmonic generation~\cite{Hsieh2010, Hsieh2011, McIver2012_2}, for review see~[\onlinecite{Ivchenko2017}].
The advantage is that some of them can be used to  excite solely the surface states even in TI materials with a high carrier density in the bulk and even at room temperature.

In this work we present a systematic study of the photogalvanic effect 
in a wide frequency range extending over two orders of magnitude from $f\approx 0.6$ to 60\,THz.
The experiments were carried out on various (Bi,Sb)Te based  3D TIs 
at room temperature. The samples, besides their composition, discriminate due to their Fermi level position or bulk carrier concentration.
For low frequency radiation and normal incidence  
the photocurrent is caused by the linear photogalvanic effect. The spectra measured reveal that they follow the Drude high-frequency conductivity varying with the radiation frequency as
$1/[1 + (2\pi f \tau)^2]$, 
where $f$ is the radiation frequency and $\tau$ is the 
scattering time of surface states carriers. 
These results are analyzed applying the microscopic theory developed in the Refs.~[\onlinecite{Olbrich2014, Plank2016}] and provide an access to the room temperature scattering times and mobilities of the surface states. 
In some samples we observed a resonance-like increase of the LPGE at high frequencies in the range from 30 to 60~THz. 
The enhancement of the LPGE current is attributed to the photoionization of Dirac fermions in the surface states to the conduction/valence band. We discuss the microscopic model of this phenomenon and show that the photocurrent is formed by a shift contribution or an asymmetric relaxation of the photo-excited electrons/holes. 
Furthermore, in this frequency range and applying oblique incidence, apart of the LPGE, we also observed a circular photogalvanic effect driven by the radiation helicity.
\section{Samples and technique}
\begin{table*}
	\begin{tabular} {p{2.5cm}||*{6}{C{2cm}|}}
		
	\rule{1pt}{0pt}	Sample 
		& Bi$_2$Te$_3$ 
		& \multicolumn{3} {c|} {Bi$_2$Te$_3$/Sb$_2$Te$_3$} 
		& \multicolumn{2} {c|} {(Bi$_{\rm 1-x}$Sb$_{\rm x}$)$_2$Te$_3$}		\\  
		& 
		& $d_{\rm ST}$ = 7.5~nm		
		& $d_{\rm ST}$ = 15~nm		
		& $d_{\rm ST}$ = 25~nm 		
		& $x$ = 0.43			
		& $x$ = 0.94	\\		
	\hline		
	
	\rule{1pt}{0pt}	$E_{\rm F}$ (meV) 		
		& 500		
		& 140	 	
		& 30		
		& -35 	 	
		& 500		
		& 7 \\    	

	\rule{1pt}{0pt}		$v_{\rm F}$ (10$^5$ m/s)		
		& 4.3		
		& 5.2	 	
		& 2.2		
		& 2.5	 	
		& 5.1		
		& 3.8  \\	

	\rule{1pt}{0pt}		$\tau$ (ps)	
		& $ > 0.25$		
		& 0.06 				
		& 0.06					
		& 0.08		 			
		& $ >  0.25$		
		& 0.04  \\		

	\rule{1pt}{0pt}	 $\mu$ (cm$^2$/Vs)
		& $ >  940$		
		& 1230	
		& 1030 
		& 1420 
		& $ >  1330$ 	
		& 8210 \cite{footnote} 
	\end{tabular}
	\caption{Samples compositions, Fermi energies $E_{\rm F}$ and Fermi velocities $v_{\rm F}$, together with scattering times $\tau$ obtained from the frequency dependencies of the linear photogalvanic effect. The scattering times are extracted with a tolerance value of 10\%. The corresponding values of the carrier mobilities $\mu$ are calculated from the scattering times and the	Fermi energies $E_{\rm F}$. Note that the latter values are measured by \textit{in-situ} ARPES and, in particular for (Bi$_{0.06}$Sb$_{0.94}$)$_2$Te$_3$ with $E_{\rm F}$ close to zero, may yield overestimated values of $\mu$, see ~[\onlinecite{footnote}].
	}
	\label{table}	
\end{table*}
\begin{figure}
	\includegraphics[width=\linewidth]{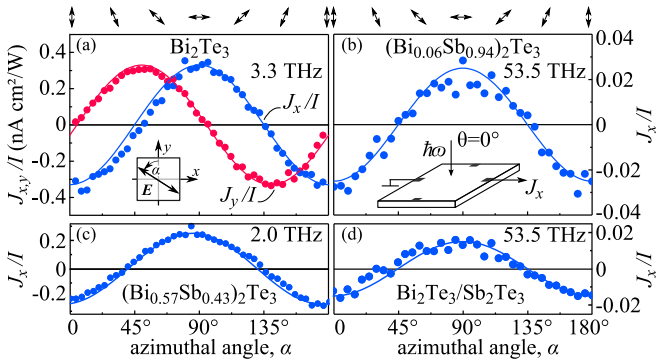}
	\caption{
		Photocurrent $J_{x,y}$ measured along $x$\,- and $y$-directions and 
		normalized on the radiation intensity $I$ in 
		(a) Bi$_2$Te$_3$,  
		(b) (Bi$_{0.06}$Sb$_{0.94}$)$_2$Te$_3$,  
		(c) (Bi$_{0.57}$Sb$_{0.43}$)$_2$Te$_3$, and  
		(d) Bi$_2$Te$_3$/Sb$_2$Te$_3$ heterostructure with Sb$_2$Te$_3$ thickness of 15~nm.  
		Solid lines show fit after Eq.~(\ref{fit}), see also Eq.~(\ref{theory}) and discussion.  Note that the polarization independent offset $D$ (${D^\prime}$), 
		being much smaller than the amplitude $A(f)$, is subtracted in these plots. The insets in panel (a) and (b) define the angle $\alpha$ and show the experimental setup. Arrows on top illustrate the polarization plane orientation for several angles $\alpha$.		
	}
	\label{figure1alpha}
\end{figure}
For this study, we used molecular beam epitaxy (MBE) 
grown (Bi$_{\rm 1-x}$Sb$_{\rm x}$)$_2$Te$_3$ based 3D TIs including a pure Bi$_2$Te$_3$ sample \cite{Plucinski2011, Kampmeier2015}, Bi$_2$Te$_3$/Sb$_2$Te$_3$ heterostructures with different thicknesses of the Sb$_2$Te$_3$ layer \cite{Eschbach2015, Lanius2016}
and (Bi$_{\rm 1-x}$Sb$_{\rm x}$)$_2$Te$_3$  ternary systems \cite{Weyrich2016}. 
The samples, grown on a (111) oriented silicon substrate, were characterized by \textit{in-situ} ARPES measurements, from which the energy dispersion and the Fermi level position were obtained.
All samples composition, Fermi velocities and energies with respect to the Dirac point are summarized in Tab.~\ref{table}. The latter varies from -35 up to 500~meV.
In the ternary systems ARPES measurements showed that in the ternary alloy with an  Antimony concentration $x=0.94$ the Fermi energy $E_{\rm F}$ lies close to the Dirac point.
Note that a reduced bulk carrier concentration  is found at $x=0.43$, see~[\onlinecite{Weyrich2016}].
In the heterostructures, where Sb$_2$Te$_3$ layers are grown on a 10~nm Bi$_2$Te$_3$ layer,  the Fermi level is tunable by varying the \textit{p}-type Sb$_2$Te$_3$  thickness $d_{\rm ST}$, see~[\onlinecite{Eschbach2015, Lanius2016}].
$X$-ray diffraction (XRD) measurements were performed to confirm the alignment of the substrate with the thin TI layer and to determine the in-plane crystallographic axes.  
With knowledge of the latter, the samples were cut along directions of 
high symmetry into 7$\times$4 mm$^2$ pieces. 
The samples were electrically contacted in the middle of the 
edges parallel to the $x$\,- and $y$-directions, see inset in Fig.~\ref{figure1alpha}(a).

To cover a wide range of frequencies numerous sources of continuous wave ($cw$) and pulsed infrared/terahertz laser radiation were applied including optically pumped molecular terahertz lasers~\cite{Ganichev1993,Schneider2004}, free electron lasers (FELBE) at the Helmholtz-Zentrum Dresden-Rossendorf
~\cite{FEL1,FEL2}, a quantum cascade laser (QCL)\cite{QCL1,QCL2} as well as $Q$-switched and transversely excited atmospheric pressure (TEA) CO$_2$ lasers~\cite{book,JETP1982,jiangPRB2010}. 
The lasers operated at single frequencies in the range from $f\approx 0.6$ to 60~THz (corresponding photon energies range from $\hbar \omega = 2.5$ to 250\,meV, where $\omega = 2\pi f$ is the angular frequency).
For the low frequency range from 0.6  to 3 THz a line-tunable 
pulsed molecular laser was used with CH$_3$F, D$_2$O and NH$_3$  
as active media~\cite{laser2,LechnerAPL2009}. The laser generated single pulses with a duration of about 100\,ns with a repetition rate of 1~Hz. The radiation intensity on the sample surface was about 10 kW/cm$^2$. 
Furthermore, low frequency measurements were performed in the range from $f=1.8$ to 10~THz with the tunable free electron laser FELBE 1 (U-27)  operating in the quasi $cw$ regime. 
The FELBE provided picosecond micro-pulses with repetition rates in the MHz range and an average power of tens of mW. 

Radiation with frequencies of about 30~THz was obtained by pulsed line-tunable $Q$-switched and  TEA CO$_2$ lasers. The Q-switched laser provided pulse durations of hundreds of nanoseconds with a peak power of about 1~kW 
and a repetition rate of about 120~Hz~\cite{jiangPRB2010}. The operation mode of the TEA CO$_2$ lasers~\cite{book} was similar to the one of the
molecular terahertz lasers.
Further lines in this range and at higher frequencies up to 60~THz were obtained 
applying the free electron laser FELBE 2 (U-100), operating in the same regime as FELBE 1 described above.
Radiation with $f=58$~THz was additionally provided by a $cw$ quantum cascade laser with a power of about 10~mW.

The peak power of the radiation was monitored, depending on the system, with Mercury Cadmium Telluride (MCT)~\cite{MCT} and photon-drag~\cite{Ganichev84p20}  detectors, 
as well as with pyroelectric  
power meters. The beam positions and 
profiles were checked with pyroelectric cameras~\cite{Ziemann2000,Drexler2012} or thermal sensitive paper. The radiation was focused onto spot sizes of about 1 to 4~mm diameter, depending on the radiation frequency. 
Experimental geometry included normal as well as oblique incidence.
In experiments at normal incidence, front and back illumination was used with corresponding angles of incidence $\theta=0$ and $180^\circ$, see inset in Fig.~\ref{figure1alpha}(b). 
The back illumination was used to ensure that the signal is caused by 
the linear photogalvanic effect, being in focus of this work, 
and to ensure that there is only a negligible contribution of the photon drag effect~\cite{Plank2016}, 
which, if present, can affect the frequency dependence of the photocurrent.
In the measurements applying oblique incident radiation, aimed at the search for the circular photogalvanic effect~\cite{Ganichev2003} in 3D TI at terahertz frequencies~\cite{McIver2012}, 
the angle of incidence $\theta$ was varied between $-40^\circ$ and $40^\circ$ with the ($yz$) plane of incidence. 
\begin{figure}
	\includegraphics[width=\linewidth]{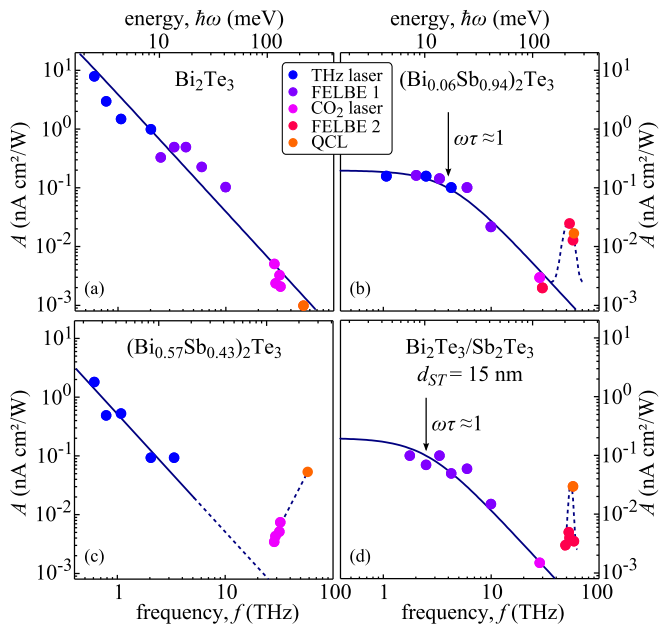}
	\caption{
		Frequency dependence of the coefficient $A$ for  
		(a) Bi$_2$Te$_3$,  
		(b) (Bi$_{0.06}$Sb$_{0.94}$)$_2$Te$_3$,  
		(c) (Bi$_{0.57}$Sb$_{0.43}$)$_2$Te$_3$, and  
		(d) Bi$_2$Te$_3$/Sb$_2$Te$_3$ heterostructure with Sb$_2$Te$_3$ thickness of 15~nm.   
		Solid line shows fit after Eq.~(\ref{fitfrq}), see also Eq.~(\ref{Drude}) and discussion. Dashed lines are guide for eye, demonstrating deviation of photocurrent amplitude from the	Drude-like behavior. 
	}
	\label{figure2frq}
\end{figure}

In the majority of the experiments, linearly polarized radiation
with an azimuthal angle $\alpha$ defining the orientation of 
the radiation electric field vector in the sample's plane 
and the $y$-axis, see inset in Fig.~\ref{figure1alpha}(a), was applied. 
The angle $\alpha$ was varied either
by rotation of half-wave plates
or a grid wire placed behind a  quarter-wave Fresnel rhomb, which was set to  provide circularly polarized radiation. 
To study the helicity dependence of the signal quarter-wave Fresnel rhombs
or plates were used.  In this geometry, the radiation helicity was 
varied as $P_{\rm circ} \propto \sin{2 \varphi}$, where the rotation angle $\varphi$ was defined 
as an angle between the laser polarization plane and the optical axis of the polarizers~\cite{15a,Kohda2012}.
Note that for $\alpha=0$  as well as for $\varphi=0$ the incident radiation was linearly polarized with electric field vector parallel to the $y$-axis.
The induced photocurrents were detected as a voltage drop 
across load resistors at room temperature. 
The signals were recorded either with GHz oscilloscopes, in case of 
the pulsed gas laser systems, or with lock-in technique, in case of the modulated quasi $cw$  radiation of  FELBE and $cw$ radiation of the QCL.
The photocurrents were measured  
in two directions, $x$ and $y$, perpendicular to each other and 
parallel to the sample edges, see insets in Fig.~\ref{figure1alpha}. 
\section{Experimental results}
A photocurrent excited by \textit{normal} incident linearly polarized radiation was detected for all used frequencies and samples. It is characterized 
by the same overall behavior: It scaled quadratically with the radiation electric field, had a response time of picoseconds or less,  and exhibited a characteristic polarization dependence.
Figure~\ref{figure1alpha}(a) presents an example of the photocurrent variation upon rotation of the radiation polarization plane obtained in Bi$_2$Te$_3$ excited with radiation frequency $f=3.3$~THz.  
The figure shows that the photocurrent scales  after 
\begin{eqnarray}
\label{fit}
    J_x(\alpha)/I &=&   A(f) s_{\rm 1} + D(f), \\
    J_y(\alpha)/I &=&  - A(f) s_{\rm 2} + D^\prime(f)\, ,  \nonumber 
\end{eqnarray}
where $s_{\rm 1} = - \cos{2 \alpha}$ and  $s_{\rm 2} = - \sin{2 \alpha}$ are the Stokes parameters of light 
defining the electric field orientation in the $x,y$ coordinate system and in a 45$^\circ$ rotated one, respectively~\cite{Stokes,BelkovJPCM}. Note that in all experiments the polarization independent offset $D(f)$ and $D^\prime(f)$ was much smaller than $A(f)$, and therefore, is out of scope of this paper. 
Figures~\ref{figure1alpha}(b)-(d) show exemplary $J_{x}(\alpha)$ measured
for further three samples including (Bi$_{\rm 1-x}$Sb$_{\rm x}$)$_2$Te$_3$ ternaries with two different Sb concentrations $x$ and one of the  
Bi$_2$Te$_3$/Sb$_2$Te$_3$ heterostructures.
Experiments with front and back illumination demonstrated that the polarization dependence itself, and sign and value of the coefficients $A(f)$ do not change. This result was the same in all samples and for all frequencies used in this work (data not shown). 
\begin{figure}
	\includegraphics[width=\linewidth]{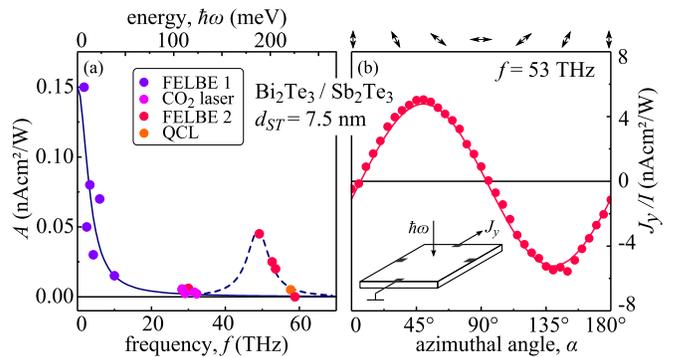}
	\caption{
		(a) Frequency dependence of coefficient $A$ of a Sb$_2$Te$_3$/Bi$_2$Te$_3$ heterostructure with $d_{\rm ST}=7.5$~nm. 
		Solid line shows fit after Eq.~(\ref{fitfrq}), see also Eq.~(\ref{Drude}) and discussion. Dashed line is guide for eye, demonstrating deviation of photocurrent amplitude from the Drude-like behavior. 
		(b) Azimuthal angle dependence of the photocurrent $J_y/I$ measured at frequency $f=53$~THz. Solid lines show fit after Eq.~(\ref{fit}), see also Eq.~(\ref{theory}) and discussion. Inset shows experimental setup. 
	}
	\label{figure3}
\end{figure}
The dependence of the coefficient $A(f)$ on the frequency is shown in Fig.~\ref{figure2frq}. 
The data reveal that in a wide range of frequencies the photocurrent decreases with the frequency increase and can be well fitted by 
\begin{equation}
\label{fitfrq}
A(f) \propto 1/[1 + (2\pi f \tau)^2]. 
\end{equation} 
\begin{figure}
	\includegraphics[width=\linewidth]{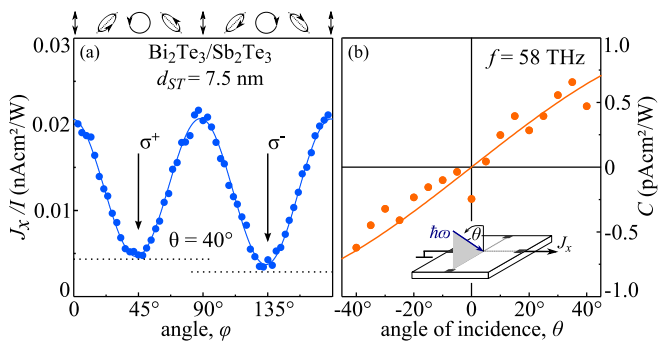}
	\caption{  
		(a) Dependence of the normalized photocurrent $J_x/I$ on the angle $\varphi$ measured in a  Sb$_2$Te$_3$/Bi$_2$Te$_3$ heterostructure with $d_{\rm ST}=7.5$~nm. 
		Data are shown for the ($yz$) plane of incidence and angle of incidence $\theta = 40^\circ$. 
		Solid line shows fit after Eq.~(\ref{fitphi}), see also Eq.~(\ref{CPGE}) and discussion. 
		Horizontal lines and downwards pointing arrows indicate photocurrent for circularly polarized radiation. 
		(b) Dependence of the circular coefficient $C$ on the angle of incidence. 
		Solid line shows fit after Eq.~(\ref{fitphi}), see also Eq.~(\ref{CPGE}) and discussion. 
		Inset shows experimental setup. 
	}
	\label{figure4}
\end{figure}
At high frequencies, however, a substantial deviation from this behavior has been detected for several samples: the photocurrent drastically increases as compared to the values expected from  Eq.(\ref{fitfrq}), reaches a maximum and then decreases, see Figs.~\ref{figure2frq}(b), (d) and~\ref{figure3}(a). Note that, while the used discrete frequencies clearly indicate the photocurrent enhancement, they do not allow a characterization of the peak with a satisfactory resolution.
As addressed above, the overall behavior at these frequencies  including its polarization and angle of incidence dependencies remained unchanged, as shown in Figs.~\ref{figure1alpha} (b), (d) and~\ref{figure3}(b) . 

For oblique incidence and linearly polarized radiation we observed the same dependence $J(\alpha)$, however, the amplitude $A(f)$ depending on sample and radiation frequency decreased or increased (data not shown, for the origin see Ref.~[\onlinecite{Plank2016}]).
Using elliptically polarized radiation, 
however, we observed that in the direction normal to 
the plane of incidence the polarization dependence was modified.  
Besides the LPGE, varying after $J_x(\varphi) = - A(f,\theta)(\cos(4\varphi)+1)/2 = A(f,\theta)s_{\rm 1}$, we observed a small but clearly pronounced 
 additional photocurrent contribution, which has opposite signs 
for right\,- and left-handed circularly polarized radiation.
The overall polarization dependence in this geometry is well described by 
\begin{equation}
\label{fitphi}
J_x(\varphi)/I =  A(f,\theta)s_{\rm 1} + C(f,\theta)s_{\rm 3} + D(f,\theta)\, ,
\end{equation} 
see Fig.~\ref{figure4}(a).
The observed circular photocurrent is proportional to the coefficient $C(f,\theta)$ and the Stokes parameter $s_{\rm 3} = \sin(2 \varphi)$ defining the radiation helicity~\cite{Stokes,BelkovJPCM}. Figure~\ref{figure4}(b) shows $C(f,\theta)$ as a function of the angle of incidence $\theta$, revealing that it  is odd in $\theta$. 
\section{Discussion}
First, we discuss the data obtained at \textit{normal} incidence. According to the symmetry analysis,  
spatially homogeneous  \textit{normal} incident radiation can result  in  
 the photogalvanic effect in the surface states as well as the photon drag effect~\cite{Olbrich2014,Plank2016}.
The corresponding  current density $\bm j$ is given by
\begin{eqnarray}
\label{theory}
    j_x &=&   (\chi +  {\cal T}  q_z) [|E_x|^2-|E_y|^2], \\
    j_y&=& - (\chi + {\cal T} q_z) [E_x E_y^*+E_yE_x^*]\,. \nonumber 
\end{eqnarray}
Here $E_{x,y}$ are in-plane projections of the radiation electric field $\bm{E}$, the factor $\chi$ is the
single linearly independent photogalvanic constant, ${\cal T}$ is the photon drag constant, $\bm q$ is the photon wavevector,  and $z$ is normal to the epilayer. 
Note that the squared brackets in the Eq.~(\ref{theory}) divided by the electric field amplitude $|E_0|^2$ represent the Stokes parameters $s_{\rm 1}$ and $s_{\rm 2}$, see~[\onlinecite{Stokes,BelkovJPCM}]. These characteristic polarization dependencies have been observed for all samples and frequencies used, see Figs.~\ref{figure1alpha} and~\ref{figure3}(b). 

The fact, that in the experiments the photocurrent amplitude $A(f)$
remains unchanged for front and back illumination excludes sizable contribution of the photon drag effect and provides a clear evidence for 
the photogalvanic effect in the surface states as a cause of the photocurrent~\cite{Olbrich2014,Plank2016}.
Indeed, the photogalvanic effect is determined only by the in-plane electric field orientation, see Eq.(\ref{theory}), and is insensitive to the radiation propagation direction. The photon drag current on the other hand, being proportional to the photon momentum ${\bm q}$, reverses its sign
at inversion of the photon wavevector $q_z$, see also Eq.(\ref{theory}), . 
Therefore, a substantial contribution of the photon drag effect should results 
either in different magnitudes $A(f)$ for front and back excitation or, if dominating,  
even in a change of  the photocurrent direction.

For small photon energies, at which Drude absorption dominates and direct optical transitions are not possible, 
the linear photogalvanic effect is
shown to be caused by the asymmetric scattering of Dirac
fermions driven back and forth by the terahertz electric field~\cite{Olbrich2014, Plank2016}.
For elastic scattering by Coulomb impurities the photogalvanic coefficient $\chi$ in Eq.~(\ref{theory}) is given by~[\onlinecite{Olbrich2014}]  
\begin{align}
	\label{chi}
	\chi = e v_{\rm F} {2\tau \over E_{\rm F}} \Xi \,\sigma(f)
\end{align}
in which  $e$ is the electric charge,  $\Xi$ the asymmetric scattering probability, $\sigma(f)$ is the high frequency (Drude) conductivity given by 
\begin{equation}
\sigma(f) = \frac{e^2 E_{\rm F} \tau}{4\pi \hbar^2 [1+(2\pi f\tau)^2]}.
\label{Drude}
\end{equation}
Equations~(\ref{chi}) and (\ref{Drude}) reveal that the amplitude of the photocurrent 
should scales after Eq.(\ref{fitfrq}). 
Our measurements performed in a wide frequency range, apart from the resonant-like increase at high frequencies observed in some samples, confirm this frequency dependence, see Figs.~\ref{figure2frq} and \ref{figure3}. 
Consequently,  the investigation of the photogalvanic effect 
allows one to analyze the Drude conductivity of the surface 
states, which provides an access to the scattering times. 
In our data the latter one can be extracted for  
(Bi$_{0.06}$Sb$_{0.94}$)$_2$Te$_3$, 
and  Bi$_2$Te$_3$/Sb$_2$Te$_3$ heterostructures 
with different thicknesses of Sb$_2$Te$_3$.
For these samples the condition $\omega\tau \approx 1$
is fulfilled, see  Figs.~\ref{figure1alpha}(b) and (d),  in the studied frequency range.  
The values of $\tau$ can be estimated from the fit functions and are summarized in Tab.~\ref{table}. 
%
%
Taking into account the Fermi level position, known from \textit{in-situ} ARPES (see Tab.~\ref{table}),
we obtained  room temperature mobilities of the Dirac states in these samples ranging from $1000$~cm$^2$/Vs up to several thousands~\cite{footnote}. 
These are of the same order of magnitude as the ones measured in transport experiments for low temperature from 2 up to 77~K, see e.g. Refs.~[\onlinecite{Weyrich2016, mobility1}]. 

The frequency dependencies of the photocurrent obtained for Bi$_2$Te$_3$ and (Bi$_{0.57}$Sb$_{0.43}$)$_2$Te$_3$ show that 
the photocurrent scales as $1/f^2$ down to the lowest frequencies used here, see Figs.~\ref{figure2frq}(a) and (c). This indicates that the value $\omega\tau$ remains substantially larger than unity. 
Consequently, the scattering times in these samples are at least by one order of magnitude larger than that for previously discussed samples.  
To determine $\tau$  in these samples further measurements with substantially lower frequencies at which $\omega\tau$ becomes less than unity are required.

Now we turn to the photocurrent enhancement observed at high frequencies in several samples. 
Our results, in particular the polarization dependence for unchanged coefficients $A$ for front and back illumination 
reveal that this photocurrent is also caused by the linear photogalvanic effect in the surface states. 
The fact that it is characterized by a non-monotonic frequency dependence and a magnitude by $10\div100$ times larger as compared to the Drude-like photocurrent clearly indicates that other types of optical transitions must be responsible for the photocurrent formation.  
This could be either direct optical transitions from the Dirac cone to the bulk states (photoionization) or interband transitions within the Dirac cone. 

Direct optical transitions induced by linearly polarized radiation can give rise to a photocurrent in systems of sufficiently low spatial symmetry (piezoelectric class)~\cite{Ivchenkobook2}. 
An example of such systems is the trigonal surface of topological insulators  studied here. 
Two contributions to the photocurrent, which are of different microscopic origin, are possible.  
First, the photocurrent can originate from the shift of electron wave packets in the real space at optical transitions (shift contribution)~\cite{Ivchenkobook2,Sturman-Fridkin-book}. 
Second, the photocurrent can emerge as a result of asymmetric relaxation of the excited electrons/holes. 
The latter mechanism of the photocurrent formation contains two stages. 
At the first stage, the optical excitation by linearly polarized radiation leads to an anisotropic distribution of carries in the momentum space which is described by the stationary correction to the electron distribution function scales as a square of the ac electric field magnitude. 
The phenomenon is known in semiconductor physics as the optical alignment of electron momenta~\cite{Olbrich2014, Mirlin1984, Golub2011}. 
At the second stage, the relaxation of the stationary correction
to the distribution function by trigonal scatterers gives rise to a directed flow of carriers, i.e., an electric current.
Similar two-step mechanisms of the photocurrent formation have been considered for the surface photocurrents in metals~\cite{Magarill1981} and bulk GaAs~\cite{Alperovich1982}, as well as for quantum well structures~\cite{Tarasenko2011}.

In the idealized pure linear dispersion model, the photoionization, which results in the depopulation of Dirac states and population of excited bulk states,  optical transitions and, consequently, related photocurrents are excited in the  range provided by $\hbar\omega > E_{\rm c} - E_{\rm F}$. 
This is because  the photoionization picture requires that the initial states of the direct optical
transitions are occupied. 
Furthermore, to excite such transitions the final states must be empty. 
Therefore these transitions take place only in a certain range of photon energies~\cite{Dantscher2017}. 
In real structures, the dispersion is more complicated and condition are not as straight forward. 
That is particularly the case in the such samples as our heterostructures combining two different materials.  
Nevertheless, the photocurrent due to photoionization must show a non-monotonic resonant-like frequency dependence as observed in experiments, see Fig.~\ref{figure2frq} and~\ref{figure3}(a). 
According to band structure calculations, the above condition is fulfilled for photon energies corresponding to the enhanced photocurrent in all three Bi$_2$Te$_3$/Sb$_2$Te$_3$ heterostructures and the 
(Bi$_{0.06}$Sb$_{0.94}$)$_2$Te$_3$ sample. Therefore, the model is relevant for the experiments. 
While this qualitatively description of the resonance seems to be appropriated, a theory is not yet developed and is crucially needed.
In the (Bi$_{0.57}$Sb$_{0.43}$)$_2$Te$_3$ sample the Fermi energy is substantially above than the edge of the conduction band, which results in the hybridization of the surface states and exclude photoionization. We attribute the observed resonance in this sample to surface photocurrents served by bulk carriers~\cite{Alperovich1982}. 

Assuming again ideal Dirac fermion bands, interband transitions become possible for photon energies larger than the double Fermi energy. 
Such systems  are characterized by a constant probability as  experimentally shown for graphene, where a value of 2.3\% of the optical absorption has been reported~\cite{grapheneabs23}.  
It seems thus unlikely that such transitions are the origin of the observed resonant-like photocurrent. 

In experiments in pure Bi$_2$Te$_3$  no deviation from the Drude-like behavior was found, see Fig.~\ref{figure2frq}(a). 
Band structure calculations for our samples demonstrate that in Bi$_2$Te$_3$ samples with $E_{\rm F} = 500$~meV direct optical transitions are prohibited in the whole range of used photon energies (up to 250~meV). 

At last but not least we discuss the circular photogalvanic effect (CPGE) detected at oblique incidence, see Fig.~\ref{figure4}.  
The observed polarization dependence as well as the dependence on the angle of incidence are in full agreement with the phenomenological theory of the CPGE in the surface states, which are characterized by the C$_{\rm 3v}$ 
point group symmetry. The corresponding photocurrent for the ($yz$) plane of incidence is given by~\cite{book,Ganichev2003,Ivchenkobook2}
\begin{equation}
\label{CPGE}
J_x^{\rm circ}(\varphi) = \gamma t_{\rm p} t_{\rm s} E_0^2 P_{\rm circ} n\sin \theta 
=  C(f, \theta) I \sin 2 \varphi \, ,
\end{equation} 
in which $\gamma$ is the CPGE constant, 
$E_0$ is the electric field amplitude in vacuum, 
$t_{\rm p}$ and $t_{\rm s}$ are transmission coefficients after 
Fresnel's formula for linear $p$- and $s$-polarizations,
and  $n$ is the refraction index.
Alike the LPGE  resonance   addressed above, we attribute the observed CPGE 
to the photoionization of the surface states~\cite{footnote2}. The microscopic mechanism, 
however, needs to take into account selective excitation of spin branches by 
circularly polarized radiation which follows from the  selection rules. 
Such processes have been previously considered for 3D TIs excited with near 
infrared radiation~\cite{McIver2012} and for 2D TIs excitation of electrons 
from helical edge states to bulk conduction band states~\cite{TItheory,Magarill2016,Dantscher2017}. 
To conclude on the mechanism responsible for the circular photocurrent observed in our 
experiments  further measurements are required, in particular, a detailed study of the  frequency dependencies of the CPGE is needed. This is a subject of future work and is out of scope of the current paper. 
\section{Conclusion}
To summarize, extensive investigation of the spectra of the photocurrent excited at normal incidence demonstrated that in very different samples and wide range of terahertz frequencies it is caused by the linear photogalvanic effect at Drude-like free carrier absorption. 
These experiments show  that spectral studies of the linear photogalvanic effect in the terahertz/microwave range allows one to 
measure the mobility of the surface states carriers. 
We emphasize that the photogalvanic effect can  only be excited in non-centrosymmetric surface states.
Thus the frequency behavior of the Drude conductivity can be studied  even at room temperature and  in materials with substantial conductance in the bulk,
where conventional surface electron transport can not be applied.
Besides the LPGE caused by Drude absorption we also observed a enhanced 
linear photogalvanic effect and the circular photogalvanic effect excited by infrared radiation, which are attributed to the ''ionization'' of
surface states at high frequencies. 
\section*{Acknowledgements} 
We thank L. E.  Golub, M. V. Durnev and S. A. Tarasenko for fruitful discussions.
 We are grateful to P. Michel and the ELBE-team for their dedicated support. 
The support from the DFG priority program SFB 1277 (project A04) and SPP1666,  
 and the Elite Network of Bavaria (K-NW-2013-247)
is gratefully acknowledged.


\begin{thebibliography}{99}
\bibitem{footnote} Note that the highest mobility value, extracted for (Bi$_{0.06}$Sb$_{0.94}$)$_2$Te$_3$ corresponds, on the first glance surprisingly, to the shortest scattering time. This is due to the very small value of the Fermi energy obtained from ARPES and in fact can be overestimated.
\bibitem{footnote2} 
	Note that in some materials, apart of topological surface states, Rashba\,-\,Dresselhaus spin\,-\,split states may also contribute to the photogalvanic effects.
\end{thebibliography}
\end{document}